\documentclass[11pt,a4paper]{article}

\usepackage{amsmath,amssymb,amsfonts}
\usepackage{graphicx}
\graphicspath{{./figures/}}
\usepackage{booktabs}
\usepackage{multirow}
\usepackage{subcaption}
\usepackage{geometry}
\usepackage{cite}
\usepackage[colorlinks=true,
            linkcolor=blue,
            citecolor=blue,
            urlcolor=blue]{hyperref}
\title{
\textbf{Reconstruction of $f(Q,T)$ Gravity from Logarithmically Corrected Ricci-Gauss-Bonnet Holographic Dark Energy}
}

\author{
Preeti Joshi$^{1}$, Ertan Gudekli$^{2}$, Antonio Pasqua$^{3}$ and
Surajit Chattopadhyay$^{1,\ast}$\\
{\small $^{1}$Department of Mathematics, Amity University Kolkata, India}\\
\small $^{2}$Department of Physics, Istanbul University, Turkey\\
\small $^{3}$Department of Physics, University of Trieste, Italy\\
\small $^{\ast}$Corresponding author: 
\href{mailto:schattopadhyay1@kol.amity.edu}{\texttt{schattopadhyay1@kol.amity.edu}}
}

\date{\today}

\begin{document}

\maketitle


\begin{abstract}
The current paper reports an investigation on the cosmological and thermodynamic behaviour of an $f(Q,T)$ modified-gravity framework in which the gravitational Lagrangian is written as $f(Q,T)=f(Q)+\lambda T$, with $Q$ denoting the non-metricity scalar and $T$ the trace of the energy-momentum tensor. The modified Friedmann equations are formulated in terms of an effective Dark Energy sector, and the corresponding equation of state and squared speed of sound are examined for a phenomenological polynomial form of $f(Q)$. A power-law background is constrained using 32 Cosmic Chronometer
measurements. The resulting background is then used to study the effective Dark Energy dynamics and its classical stability in the redshift range considered. In this study, we further construct a generalized Ricci-Gauss-Bonnet holographic Dark Energy model with logarithmic entropy corrections based on the Nojiri-Odintsov prescription and establish a correspondence between its energy density and the effective Dark Energy density of the $f(Q,T)$ framework. The full function $f(Q,T)$ is then given by adding the trace contribution $\lambda T$. Finally, the thermodynamic behaviour of the reconstructed model is studied at the apparent horizon with the Nojiri-Odintsov entropy motivated by \textit{Phys. Rev. D} \textbf{105}, 044042 (2022) and the Gibbs relation. The entropy evolution obtained from the compact background description is also found to be consistent with that obtained from the reconstructed effective-fluid formulation, with only small numerical residuals. These results provide a consistent framework for examining the connection between generalized holographic dark energy, reconstructed $f(Q,T)$ gravity, and cosmic thermodynamics.
\\
\textbf{keywords} {$f(Q,T)$ gravity; \and holographic dark energy; \and Nojiri-Odintsov entropy; \and Ricci-Gauss-Bonnet cutoff; \and cosmological reconstruction; \and observational constraints.}

\end{abstract}
 \section{Introduction}
The accelerating expansion of the universe is a significant topic in cosmology \cite{livio2001accelerating,liddle2001acceleration,caldwell2009physics,kragh2006conceptions}. This acceleration cannot be explained by matter or radiation alone, indicating the presence of dark energy in the universe (see references \cite{livio2001accelerating,liddle2001acceleration,caldwell2009physics,kragh2006conceptions,copeland2006dynamics,oks2021brief,peebles2003cosmological,mortonson2013dark} ). Researchers have proposed various models and ideas to investigate the genesis and nature of the unknown energy \cite{copeland2006dynamics}. One essential and fascinating concept is Holographic Dark Energy (HDE) \cite{Nojiri2021,NojiriOdintsov2017,Ualikhanova2024,Nojiri2022,Nojiri2020,Elizalde2005,Nojiri2006}. This model is based on the holographic principle \cite{Bousso2000,Bousso2002}, which originates from quantum gravity. The holographic principle states that the information of a region of space can be represented by the information on its boundary surface. This notion can also be applied to the entire universe \cite{Bousso2000, Bousso2002}. In the HDE model, the density of dark energy is inversely proportional to the square of a given length, known as the IR cutoff. This length might be associated with the Hubble horizon, the future event horizon or other cosmic scales. This model connects the small-scale (ultraviolet) and large-scale (infrared) constraints in quantum theory, giving a dynamical version of dark energy. There are plethoras of studies in the literature that have shown that the HDE can suitably explain the accelerated expansion in the late time of the universe and has certain connections with the quantum gravity theories  \cite{Luciano2025,Vinutha2022}. As a result, it is considered one of the best models for describing current cosmic acceleration, surpassing the simple cosmological constant model.

A useful paradigm for explaining the  acceleration of the universe in both early and late times is holographic cosmology \cite{Bousso2002}. This method uses the holographic principle to relate the cosmic energy density to an infrared cutoff scale associated with degrees of freedom in quantum gravity \cite{Guberina2005}. In this context, holographic inflation was proposed, in which holographic energy density, rather than an explicit scalar field, drives the early accelerated growth \cite{Nojiri:2019kkp}. The selection of the IR cutoff has a significant impact on the cosmic behaviour of holographic dark energy \cite{Li2004}. In the works of \cite{Gao2009,Saridakis2018}, curvature-based extensions have been proposed to solve the drawbacks of horizon-based cutoffs. In these extensions, the energy density is built using geometric invariants such the Ricci scalar $R$ and the Gauss-Bonnet term $G$. Better agreement with observational data is demonstrated by these models \cite{Li2004}. Cosmic acceleration can also be consistently described by other methods, such as holographic bounce models \cite{Nojiri2019Bounce} and covariant formulations that combine holographic inflation with dark energy.

Holographic inflation is a holographic depiction of the inflationary epoch, where the early accelerated expansion is driven by holographic energy density instead of an ad hoc scalar field \cite{Nojiri:2019kkp}. This concept was developed in the study of \cite{Nojiri2022} to a unified covariant formulation, that provided a consistent description of cosmic acceleration throughout the  evolution of the universe \cite{Nojiri2022}. These advancements have prompted the introduction of holographic cosmology into the framework of modified gravity like $f(G)$, $f(Q)$, and $f(T)$ \cite{Aghamohammadi2014,Aly2019,Cai2016,DeFelice2009,Bamba2017,Odintsov2016,Mandal2020,Anagnostopoulos2021,Xu2019}. Modified Gauss-Bonnet gravity has been extensively investigated as a geometrical alternative for dark energy and late-time cosmic acceleration and literatures include \cite{NojiriOdintsov2005,Cognola2006}. More recently, well-defined $f(Q)$ gravity has been developed in connection with the reconstruction of FLRW spacetime and the possible unification of inflation with the late-time Dark Energy epoch \cite{NojiriOdintsov2024fQ}. While Gudekli et al.~\cite{Gudekli2015} investigated trace anomaly-induced inflation in $f(T)$ gravity, Esmakhanova et al.~\cite{Esmakhanova2011} investigated dark energy behaviour in integrable and non-integrable FRW models. Using Noether symmetry with fermionic fields, Myrzakulov et al.~\cite{Myrzakulov2019} advanced $f(T)$ cosmology, and the metric-affine extension was subsequently developed in \cite{Myrzakulov2021}. The $f(Q)$ symmetric teleparallel gravity has gained considerable attention in recent times \cite{jimenez2018coincident,PhysRevD.101.103507}. Inspired by $f(R)$\cite{buchdahl1970non, capozziello2002curvature} and $f(T)$ gravity\cite{PhysRevD.84.043527,farrugia2016stability,PhysRevD.97.044008}, this framework generalizes symmetric teleparallel gravity\cite{nester1999symmetricteleparallelgeneralrelativity} by substituting an arbitrary function $f(Q)$. Additionally, holographic dark energy models can be reconstructed using geometric quantities \cite{Mandal2020,Anagnostopoulos2021,Xu2019}. Furthermore, second-order field equations are involved in $f(T)$ and $f(Q)$ theories. As a result, at the background level, $f(Q)$ gravity has produced a number of important cosmological scenarios\cite{dialektopoulos2019noether,bajardi2020bouncing,PhysRevD.103.063505,flathmann2021post}. An arbitrary function of the non-metricity $Q$ and the trace of the matter-energy momentum tensor $T$ determines the Lagrangian density in the recently suggested $f(Q, T)$  gravity theory \cite{Xu2019}, which is an extension of symmetric teleparallel gravity. As with $f(R, T)$ gravity \cite{Harko2011}, the coupling between $Q$ and $T$ leads to the non-conservation of the energy-momentum tensor, which implies a substantial thermodynamic change of the Universe \cite{harko2018extensions}. Recent developments have also explored fermion localization and brane dynamics in $f(Q,\mathcal{T})$ gravity with noncanonical scalar-field dynamics \cite{MoreiraDongSaridakis2026}.

In cosmology, thermodynamics is a crucial area of study. To determine whether the laws of thermodynamics are applicable in the presence of dark energy, one treats the universe as a thermodynamic system. This helps in our comprehension of the consistency and stability of many cosmological hypotheses. Different dark energy models may also be constrained by the generalized second law of thermodynamics (GSLT). As a result, cosmic evolution may now be studied with the help of the thermodynamic description of the universe. In particular, modified gravity models have extensively examined the relationship between horizon thermodynamics and gravitational dynamics. The second law holds true throughout the cosmic evolution, as demonstrated by Odintsov et al.~\cite{Odintsov2025}. The thermodynamic behaviour of a world dominated by dark energy, including the potential for future singularities, was explored by Nojiri and Odintsov~\cite{Nojiri2004}. Later, Nojiri et al.~\cite{Nojiri20201} demonstrated that non-extensive thermodynamics and generalized fluid models are equivalent. Using Barrow entropy, Saridakis and associates\cite{Saridakis2021,Saridakis2020,Saridakis2020Thermo} investigated the GSLT further and demonstrated how variations from Bekenstein-Hawking entropy can impact the universe's late-time thermodynamic behaviour. Additional incentive comes from recent advances in gravitational thermodynamics. Specifically, the cosmological behaviour is improved by curvature-based extensions of holographic dark energy including the Ricci and Gauss-Bonnet invariants $R$ and $G$ \cite{Saridakis2018,NojiriOdintsov2017}

The discovery of the late-time accelerated expansion of the Universe is a milestone of modern cosmology \cite{Perlmutter1999,Riess1998}. The first direct evidence for this phenomenon came from independent observations of Type Ia supernovae, followed by confirmation of time-dilation measurements of spectral evolution \cite{Foley2005}. Further studies of the intrinsic diversity among these supernovae only strengthen the case for strong systematic control for precision cosmology \cite{Wang2012}. These combined results showed that the Universe is dominated by an exotic component with negative pressure, generally known as dark energy \cite{DES2016,Copeland2006,Frieman2008,Peebles2003,Li2013,Yoo2012,Sahni2006,Oks2021}.These observational breakthroughs have shifted the focus to discriminating between simple cosmological constant and more complex dynamical dark energy models \cite{Copeland2006,Peebles2003,Li2013,Yoo2012,Sahni2006}. Large scale surveys such as the Dark Energy Survey (DES) have improved our constraints on the expansion history and structure growth \cite{DES2016} but the theoretical interpretation is still under debate. More recent analyzes by Alfano et al. \cite{Alfano2025} show that DESI data do not exclude dynamical dark energy, but the $\Lambda$CDM model is still statistically preferred due to its parsimony. Modified gravity has emerged as a robust framework to explain cosmic acceleration without invoking an explicit dark energy component. By offering rich phenomenology across inflationary, bouncing, and late-time regimes \cite{Koyama2016,Joyce2016,Clifton2012,NojiriOikonomou2017,Bamba2015}, these theories naturally account for large-scale observations while maintaining consistency with local constraints \cite{Koyama2016,Clifton2012}.

Recent research has further expanded this scope to include generalized entropy, non-zero torsion, and Aether scalar-tensor theories, demonstrating that modified geometries can satisfy stringent thermodynamic and observational requirements \cite{Rani2026,Alruwaili2026,Azhar2025,Usman2025}. Parallel to these geometric modifications, the holographic principle provides an alternative approach where dark energy is interpreted through the entropy-area bound \cite{Dubey2021,Wang2023,DelCampo2011}. This paradigm has evolved through the introduction of generalized infrared (IR) cutoffs constructed from curvature invariants \cite{NojiriOdintsov2017}. By incorporating modified entropy relations--such as the Nojiri-Odintsov entropy--these models capture high-energy gravitational effects that standard vacuum fluctuations often fail to describe \cite{ElizaldeNojiriOdintsovWang2005,NojiriOdintsovPaul2022,Jawad2022b}. These developments motivate our choice of Ricci-Gauss-Bonnet (RGB) holographic models, which integrate higher-order curvature terms to refine the description of cosmic expansion \cite{Dabash2025,Bueno2019,AguilarGutierrez2023}. In the present work, we investigate the cosmological implications of a generalized Ricci-Gauss-Bonnet Holographic Dark Energy (GRGB-HDE) model within the framework of $f(Q,T)$ gravity, where $Q$ represents the non-metricity scalar and $T$ denotes the trace of the energy-momentum tensor. This geometric framework provides a compelling foundation, as the gravitational interaction is driven by the non-metricity of the spacetime connection, avoiding the explicit curvature of General Relativity. 

The main novelty of the present work lies in constructing a direct reconstruction link between logarithmically corrected Ricci--Gauss--Bonnet holographic Dark Energy and the geometric sector of $f(Q,T)$ gravity. Unlike phenomenological choices of $f(Q)$, the reconstructed $f(Q)$ function is obtained directly from the correspondence $\rho_{\rm DE}=\rho_{\rm GRGB}$ without imposing a prescribed functional form. The reconstructed model is further constrained by Cosmic Chronometer observations and examined through the effective equation of state, squared speed of sound, and viability of the geometric sector. In addition, the thermodynamic consistency of the reconstruction is tested using the Nojiri--Odintsov logarithmically corrected entropy and the generalized second law, including a comparison between the compact background and reconstructed effective-fluid descriptions. Finally, the reconstructed geometric sector is discussed in the context of the
recently established generalized-entropic--$f(Q)$ correspondence, with the $\lambda T$ term providing an extension to a matter--geometry coupled $f(Q,T)$ framework. It may be noted that here, by the compact background description we mean the direct evaluation of the thermodynamic quantities from the background cosmological variables $H$, $\dot H$, $\rho_{\rm GRGB}$ and the corresponding horizon entropy, without explicitly reconstructing the effective $f(Q,T)$ related variables.

Rest of the paper is organized as follows: In section~\ref{sec:fQT}, the theoretical framework of $f(Q,T)$ gravity and the corresponding modified Friedmann equations in an effective dark energy form are presented. We also examine a phenomenological version of $f(Q)$ and use the squared speed of sound to study the classical stability and the effective equation of state parameter. The 32 Cosmic Chronometer observations are used to constrain the power-law cosmic background. In section~\ref{sec:GRGBHDE}, we discuss the generalized Ricci-Gauss-Bonnet holographic dark energy model with the Nojiri-Odintsov logarithmic entropy corrections and show that it corresponds to the effective dark energy sector of $f(Q,T)$ gravity. we then reconstruct the geometric function $f(Q)$ corresponding to the GRGB-HDE density to obtain the full modified-gravity function $f(Q,T)=f(Q)+\lambda T$. Section~\ref{sec:Thermodynamics} investigates the thermodynamic behaviour of the reconstructed model at the apparent horizon using the Nojiri-Odintsov entropy and the Gibbs relation, as well as the generalized second law of thermodynamics. We also compare the entropy evolution derived from the reconstructed effective-fluid formulation with that derived from the compact background description. Finally, in section~\ref{sec:conclusion} we summarize the key findings and explore the physical implications of the reconstructed $f(Q,T)$ framework. 

\section{\texorpdfstring{Essentials of $f(Q,T)$ Gravity}{Essentials of f(Q,T) Gravity}}
\label{sec:fQT}

The $f(Q,T)$ gravity represents an extension of symmetric teleparallel gravity in which the gravitational Lagrangian is taken to be an arbitrary function of the non-metricity scalar $Q$ and the trace of the energy-momentum tensor $T$ \cite{Xu2019}. A fully covariant formulation of the theory and a corrected energy-momentum balance equation were subsequently developed, providing an important refinement of the original formulation \cite{Loo2023}. The inclusion of the trace $T$ introduces a direct coupling between matter and geometry, which may provide a possible explanation for the late-time accelerated expansion of the Universe without invoking an explicit cosmological constant. Since its introduction, $f(Q,T)$ gravity has attracted considerable attention in cosmological studies due to its rich phenomenology and promising observational implications \cite{arora2020energy,Arora2020PDU}. Furthermore, various investigations based on energy conditions, observational constraints and Weyl-type extensions have shown that $f(Q,T)$ gravity can provide a viable framework for describing the present cosmic evolution and may serve as an interesting alternative to General Relativity \cite{Arora2021,Alfedeel2024}. Motivated by these developments, we consider in the present work the simple matter-geometry coupling $f(Q,T)=f(Q)+\lambda T$, which allows us to investigate the effects of generalized Ricci-Gauss-Bonnet holographic dark energy in a mathematically tractable manner. Therefore, in the present work, we consider the following functional form
\begin{equation}
f(Q,T) = f(Q) + \lambda T,
\label{eq:fQTmodel}
\end{equation}
where $\lambda$ is a constant coupling parameter governing the interaction between matter and geometry.

The total action of the theory, following the $f(Q,T)$ formulation of Xu et al.~\cite{Xu2019}, is given by
\begin{equation}
S = \int \left[ \frac{1}{16\pi} \Big(f(Q) + \lambda T\Big) + \mathcal{L}_{m} \right] \sqrt{-g} \, \mathrm{d}^{4}x,
\label{eq:action}
\end{equation}
where $\mathcal{L}_{m}$ denotes the matter Lagrangian density and $g$ is the determinant of the metric tensor $g_{\mu\nu}$.

We assume a spatially flat Friedmann-Lema\^{i}tre-Robertson-Walker (FLRW) spacetime described by the line element
\begin{equation}
\mathrm{d}s^{2} = -\mathrm{d}t^{2} + a^{2}(t) \left( \mathrm{d}x^{2} + \mathrm{d}y^{2} + \mathrm{d}z^{2} \right),
\label{eq:metric}
\end{equation}
where $a(t)$ is the cosmological scale factor. The expansion rate is characterized by the Hubble parameter
\begin{equation}
H = \frac{\dot{a}}{a}.
\label{eq:Hubble}
\end{equation}
For this geometric background, the non-metricity scalar simplifies to
\begin{equation}
Q = 6H^{2}.
\label{eq:Q}
\end{equation}
We adopt the sign convention used in the original $f(Q,T)$ formulation of Xu et al.~\cite{Xu2019}. The cosmic matter content is modeled as a perfect fluid, whose energy-momentum tensor is defined as
\begin{equation}
T_{\mu\nu} = (\rho + p)u_{\mu}u_{\nu} + pg_{\mu\nu},
\label{eq:Tmunu}
\end{equation}
where $\rho$, $p$, and $u_{\mu}$ denote the energy density, isotropic pressure, and four-velocity of the fluid, respectively. The trace of the energy-momentum tensor is given by
\begin{equation}
T = g^{\mu\nu}T_{\mu\nu} = -\rho + 3p.
\label{eq:Ttrace}
\end{equation}

Here, $T$ denotes the trace of the matter energy-momentum tensor and provides the explicit matter-geometry coupling through the $\lambda T$ term in $f(Q,T)=f(Q)+\lambda T$. For convenience, we define the partial derivatives of the functional $f(Q)$ with respect to $Q$ as
\begin{equation}
f_Q \equiv \frac{\mathrm{d}f}{\mathrm{d}Q}, \qquad f_{QQ} \equiv \frac{\mathrm{d}^{2}f}{\mathrm{d}Q^{2}}.
\label{eq:fQdef}
\end{equation}

Varying the action \eqref{eq:action} with respect to the metric tensor yields the modified gravitational field equations. For the background FLRW geometry, these equations reduce to the generalized Friedmann equations\cite{arora2020energy}:
\begin{equation}
3H^{2}=\frac{1}{2f_Q}\left[\frac{f}{2}-8\pi\rho-\lambda(\rho+p)\right],
\label{eq:friedmann1}
\end{equation}
and
\begin{equation}
2\dot{H}+3H^{2} =\frac{f-8H\dot{f}_{Q}+16\pi\rho +2\lambda(\rho+p)+32\pi p}{4f_{Q}},
\label{eq:friedmann2}
\end{equation},

where the overdot denotes a derivative with respect to cosmic time $t$. We use the same FLRW convention and matter variation as in the standard $f(Q,T)$ formulation, for which $f_T=\lambda$ and $Q=6H^2$ \cite{Xu2019}.

Differentiating Eq.~\eqref{eq:Q} with respect to time yields
\begin{equation}
\dot{Q} = 12H\dot{H},
\label{eq:Qdot}
\end{equation}
which allows us to express the time derivative of $f_Q$ as
\begin{equation}
\dot{f}_Q = f_{QQ}\dot{Q} = 12H\dot{H} f_{QQ}.
\label{eq:fQdot}
\end{equation}
Substituting Eq.~\eqref{eq:fQdot} into Eq.~\eqref{eq:friedmann2} yields the explicit form of the second Friedmann equation:
\begin{equation}
2\dot{H}+3H^2 =\frac{f-96H^2\dot{H}f_{QQ}+16\pi\rho+2\lambda(\rho+p)+32\pi p}{4f_Q}
\label{eq:friedmann2final}
\end{equation}

These modified cosmological equations can be cast into an intuitive, Einstein-like form by isolating the effective dark energy contributions:
\begin{align}
3H^{2} &= 8\pi \left( \rho + \rho_{\mathrm{DE}} \right), \label{eq:einstein1} \\
2\dot{H} + 3H^{2} &= -8\pi \left( p + p_{\mathrm{DE}} \right), \label{eq:einstein2}
\end{align}
where the effective dark energy density $\rho_{\mathrm{DE}}$ and the effective pressure $p_{\mathrm{DE}}$ emerging from the modified gravity sector are defined respectively as
\begin{equation}
\rho_{\mathrm{DE}}=\frac{1}{32\pi f_Q}\left[ f -16\pi\rho-2\lambda(\rho+p)-32\pi f_Q\rho\right],
\label{eq:rhoDE}
 \end{equation}
\begin{equation}
p_{\mathrm{DE}}=\frac{1}{32\pi f_Q}\left[-f+8H\dot{f}_Q-16\pi\rho-2\lambda(\rho+p)-32\pi(f_Q+1)p\right].
\label{eq:pDE}
\end{equation}
It should be emphasized that the effective dark energy density and pressure defined in Eq.~\eqref{eq:rhoDE} and \eqref{eq:pDE} come from rewriting the modified Friedmann equations in the forms like Einstein. Consequently these two equations represent effective quantities involving both the geometric modifications and matter geometry coupling, rather than an independent dark energy fluid. Consequently, the effective dark energy equation-of-state (EoS) parameter is given by
\begin{equation}
w_{\mathrm{DE}}=\frac{-f+8H\dot{f}_Q-16\pi\rho-2\lambda(\rho+p)-32\pi(f_Q+1)p}{f-16\pi\rho-2\lambda(\rho+p)-32\pi f_Q\rho}
\label{eq:wDE}
\end{equation}

The effective EoS parameter in Eq.~\eqref{eq:wDE} is determined by the ratio of the effective pressure and density obtained from the modified Friedmann equations. The limit $f(Q)=Q$ should be interpreted together with the corresponding convention for the non-metricity scalar and the matter-geometry coupling; it is therefore not used as a separate phenomenological Dark Energy model in the following analysis. However, presence of the term $\lambda (\rho +p)$ shows that the interaction between matter and geometry modifies the effective equation of state. Therefore Eq.~\eqref{eq:wDE} , involving $H, \dot{H}$ and $Q$ serve as a fundamental diagnostic for investigating the cosmological evolution

\subsection{\texorpdfstring{Polynomial Parameterization of the $f(Q)$ Geometric Sector}{Polynomial Parameterization of the f(Q) Geometric Sector}}

In this subsection, to investigate the cosmological behaviour and stability properties of the geometric sector, we first consider a phenomenological polynomial parameterization of $f(Q)$, given by
\begin{equation}
f(Q) = Q + \alpha Q^{m} + \beta Q^{q}, \qquad m > 1, \qquad q > m,
\label{eq:polyfQ}
\end{equation}
where $\alpha$ and $\beta$ are constant model parameters. The linear term represents the symmetric teleparallel equivalent of General Relativity (STEGR), while the higher-order terms serve as power-law corrections that characterize deviations from the standard cosmological paradigm \cite{jimenez2018coincident,PhysRevD.101.103507}. Here, this parameterization is considered as a phenomenological representation of the geometric contribution. It should be noted that it is not imposed as an ansatz in the subsequent GRGB-HDE reconstruction. In particular, the reconstruction developed here determines $f(Q)$ from the correspondence of GRGB-HDE through the resulting differential equation, where no polynomial form would be assumed. Thus, the polynomial analysis presented here provides an investigation of the cosmological behaviour and stability associated with representative nonlinear $f(Q)$ corrections. Whereas, the reconstruction section to be reported later determines the gravitational function directly from the holographic correspondence.

Differentiating Eq.~\eqref{eq:polyfQ} with respect to the non-metricity scalar $Q$ yields
\begin{equation}
f_Q = 1 + \alpha m Q^{m-1} + \beta q Q^{q-1},
\label{eq:fQ}
\end{equation}
and
\begin{equation}
f_{QQ} = \alpha m(m-1) Q^{m-2} + \beta q(q-1) Q^{q-2}.
\label{eq:fQQ}
\end{equation}
Substituting Eqs.~\eqref{eq:fQ} and \eqref{eq:fQQ}, together with the polynomial form of $f(Q)$ given in Eq.~\eqref{eq:polyfQ}, into the effective equation-of-state parameter defined in Eq.~\eqref{eq:wDE}, we obtain the following explicit expression:

\begin{equation}
\begin{aligned}
w_{\mathrm{DE}} = 
\left[ Q+\alpha Q^{m}+\beta Q^{q} - 16\pi\rho - 2\lambda(\rho+p) - 32\pi\rho \left(1+\alpha mQ^{m-1} + \beta qQ^{q-1}\right) \right]^{-1} \\
\times \Big[ -\left(Q+\alpha Q^{m}+\beta Q^{q}\right) 
+ 96H^{2}\dot{H} \left( \alpha m(m-1)Q^{m-2} + \beta q(q-1)Q^{q-2} \right) \\
- 16\pi\rho - 2\lambda(\rho+p) 
- 32\pi p \left( 2+\alpha mQ^{m-1} + \beta qQ^{q-1} \right) \Big].
\end{aligned}
\label{eq:wDE_poly}
\end{equation}
Let us first consider the power law scale factor, 
\begin{equation}
    a(t)= a_0 t^n
    \label{eq:power law}
\end{equation} where $n>1$ is the expansion index that describes the  dynamics of the universe and $a_{0}$ is a positive constant that represents the  present value of the scale factor. The Hubble parameter and its temporal derivatives for this parameterization are provided by, 
\begin{equation}
H = \frac{\dot{a}}{a} = \frac{n}{t}, 
\label{eq:HubblePowerLaw}
\end{equation}
\begin{equation}
\dot{H} = -\frac{n}{t^{2}}, 
\qquad 
\ddot{H} = \frac{2n}{t^{3}}.
\label{eq:Hdot}
\end{equation}

Before proceeding further, we would like to mention that we have considered the power law scale factor as a phenomenological background ansatz for the reconstruction. This choice leads to a mathematically tractable description for the cosmic expansion such that $H=n/t$ and $\dot{H}=-n/t^{2}$. In this way, the quantities of the curvature and the non-metricity scalar entering in the GRGB-HDE and $f(Q,T)$ sectors can be derived in a relatively simple form. The exponent $n$ has a direct relation with the expansion regime, $n>1$ corresponds to accelerated power-law expansion and $0<n<1$ to decelerated one. The power law ansatz is used here as a controlled reconstruction scheme and not as an assumption about the exact power law nature of the real cosmic expansion at all epochs. Then, the viability of the reconstructed model obtained is checked through observational and thermodynamic considerations.

For the power-law scale factor mentioned above, the corresponding non-metricity scalar becomes, $Q= \frac{6n^2}{t^2}$. By substituting the expression of the non-metricity scalar,
\[
Q=\frac{6n^2}{t^2},
\]
the polynomial model and its first and second derivatives can be expressed in terms of the cosmic time $t$ as
\begin{equation}
\begin{aligned}
f(Q)
=6n^{2}t^{-2}
+\alpha(6n^{2})^{m}t^{-2m}
+\beta(6n^{2})^{q}t^{-2q}.
\end{aligned}
\label{eq:fQ_powerlaw}
\end{equation}

\begin{equation}
f_Q
=
1+\alpha m(6n^2)^{m-1}t^{-2m+2}
+\beta q(6n^2)^{q-1}t^{-2q+2},
\label{eq:fQ_t}
\end{equation}

\begin{equation}
f_{QQ}=\alpha m(m-1)(6n^2)^{m-2}t^{-2m+4} +\beta q(q-1)(6n^2)^{q-2}t^{-2q+4}.
\label{eq:fQQ_t}
\end{equation}

and

\begin{equation}
f_{QQQ} = \alpha m (m - 1)(m - 2) (6n^2)^{m - 3} t^{-2m + 6} + \beta q (q - 1)(q - 2) (6n^2)^{q - 3} t^{-2q + 6} .
\label{eq:fQQQ_t}
\end{equation}

Substituting the power-law background $H=n/t$, $\dot{H}=-n/t^{2}$ and $Q=6n^{2}/t^{2}$, together with Eqs.~\eqref{eq:polyfQ}-\eqref{eq:fQQ}, into Eq.~\eqref{eq:wDE_poly}, we obtain the effective EoS parameter as a function of cosmic time $t$ as follows:

\begin{equation}
w_{\mathrm{DE}}(t) = \frac{
  \begin{aligned}
    &-\!\left[ \frac{6n^2}{t^2} + \alpha\left(\frac{6n^2}{t^2}\right)^m + \beta\left(\frac{6n^2}{t^2}\right)^q \right] \\
    &-\frac{96n^3}{t^4} \left[ \alpha m(m-1)\left(\frac{6n^2}{t^2}\right)^{m-2} + \beta q(q-1)\left(\frac{6n^2}{t^2}\right)^{q-2} \right] \\
    &- 16\pi\rho - 2\lambda(\rho+p) - 32\pi p \left[ 2 + \alpha m\left(\frac{6n^2}{t^2}\right)^{m-1} + \beta q\left(\frac{6n^2}{t^2}\right)^{q-1} \right]
  \end{aligned}
}{
  \begin{aligned}
    &\frac{6n^2}{t^2} + \alpha\left(\frac{6n^2}{t^2}\right)^m + \beta\left(\frac{6n^2}{t^2}\right)^q - 16\pi\rho - 2\lambda(\rho+p) \\
    &- 32\pi\rho \left[ 1 + \alpha m\left(\frac{6n^2}{t^2}\right)^{m-1} + \beta q\left(\frac{6n^2}{t^2}\right)^{q-1} \right]
  \end{aligned}
}.
\label{eq:wDE_t}
\end{equation}

Now, For convenience, the numerators appearing in the effective dark energy density and pressure in Eqs.~\eqref{eq:rhoDE} and \eqref{eq:pDE} are denoted by
\begin{align}
\mathcal{N} &= f - 16\pi\rho - 2\lambda(\rho+p) - 32\pi f_Q\rho, \label{eq:N_def} \\[1ex]
\mathcal{P} &= -f + 8H\dot{f}_Q - 16\pi\rho - 2\lambda(\rho+p) - 32\pi(f_Q+1)p. \label{eq:P_def}
\end{align}
Thus, $\rho_{\mathrm{DE}} = \mathcal{N} / (32\pi f_Q)$ and $p_{\mathrm{DE}} = \mathcal{P} / (32\pi f_Q)$, and the squared speed of sound is defined as
\begin{equation}
v_s^2=\frac{\dot{p}_{\rm DE}}{\dot{\rho}_{\rm DE}}=\frac{f_Q\dot{\mathcal{P}}-\mathcal{P}f_{QQ}\dot{Q}}{f_Q\dot{\mathcal{N}}-\mathcal{N}f_{QQ}\dot{Q}}.
\label{eq:vs2}
\end{equation}

\begin{figure*}[t]
    \centering

    \begin{minipage}{0.47\textwidth}
        \centering
        \includegraphics[width=\textwidth]{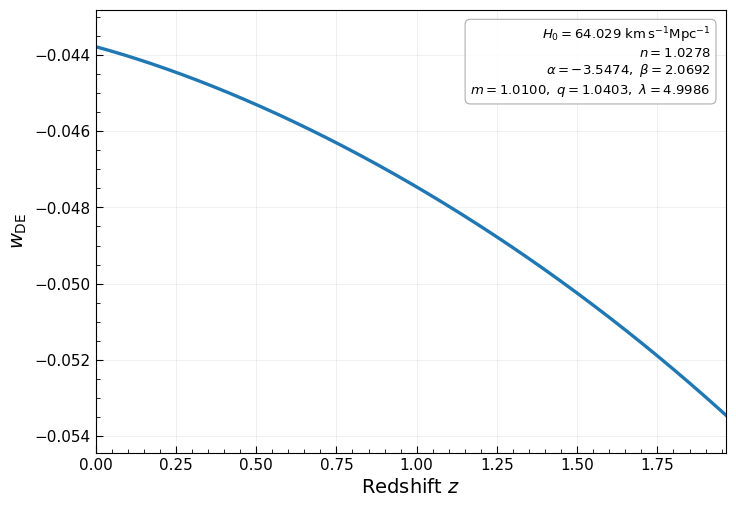}
        \captionof{figure}{Evolution of the Dark Energy equation-of-state
        parameter $w_{\rm DE}$ as a function of redshift $z$ for the
        reconstructed $f(Q)$ model. The fixed background parameters are
        $H_0=64.0295~\mathrm{kms^{-1}\,Mpc^{-1}}$ and
        $n=1.02782$, while the viable parameter set is
        $\alpha=-3.5474$, $\beta=2.0692$, $m=1.0100$,
        $q=1.0403$, and $\lambda=4.9986$.}
        \label{fig:wDE}
    \end{minipage}
    \hfill
    \begin{minipage}{0.47\textwidth}
        \centering
        \includegraphics[width=\textwidth]{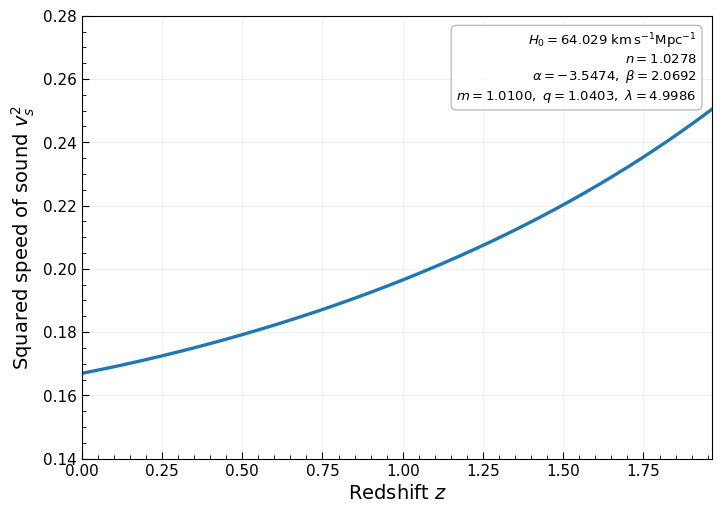}
        \captionof{figure}{Evolution of the squared speed of sound
        $v_s^2$ as a function of redshift $z$ for the reconstructed
        $f(Q)$ model. The fixed background parameters are
        $H_0=64.0295~\mathrm{kms^{-1}\,Mpc^{-1}}$ and
        $n=1.02782$, while the viable parameter set is
        $\alpha=-3.5474$, $\beta=2.0692$, $m=1.0100$,
        $q=1.0403$, and $\lambda=4.9986$.}
        \label{fig:vs2}
    \end{minipage}

\end{figure*}

For the power-law cosmological background $a(t)=a_{0}t^{n}$, the corresponding Hubble parameter and the non-metricity scalar are $H=n/t$ and $Q=6H^{2}$, respectively. We assume that the matter sector is pressureless ($p=0$) and that the matter density evolves as $\rho=\rho_{0}a^{-3}$, with $\rho_{0}$ being the present matter density of the Universe. Plugging these relations along with the polynomial form $f(Q)=Q+\alpha Q^{m}+\beta Q^{q}$ into Eq.~(\ref{eq:vs2}), the squared speed of sound is obtained numerically in the relevant redshift range. Fig.~\ref{fig:vs2} shows the resulting evolution of $v_{s}^{2}$ which is then used to investigate the classical stability of the reconstructed cosmological model.

\begin{figure}[h!]
\centering

\begin{subfigure}[t]{0.95\textwidth}
    \centering
    \includegraphics[width=0.80\textwidth]{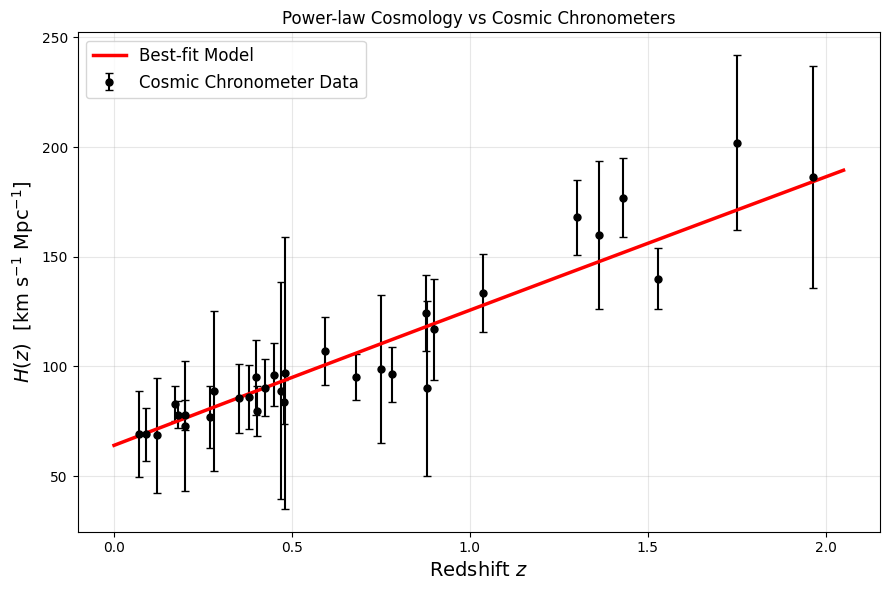}
    \caption{Comparison of the best-fit Hubble parameter with the 32 Cosmic Chronometer measurements.}
    \label{fig:HzCC}
\end{subfigure}

\vspace{0.4cm}

\begin{subfigure}[t]{0.48\textwidth}
    \centering
    \includegraphics[width=\textwidth]{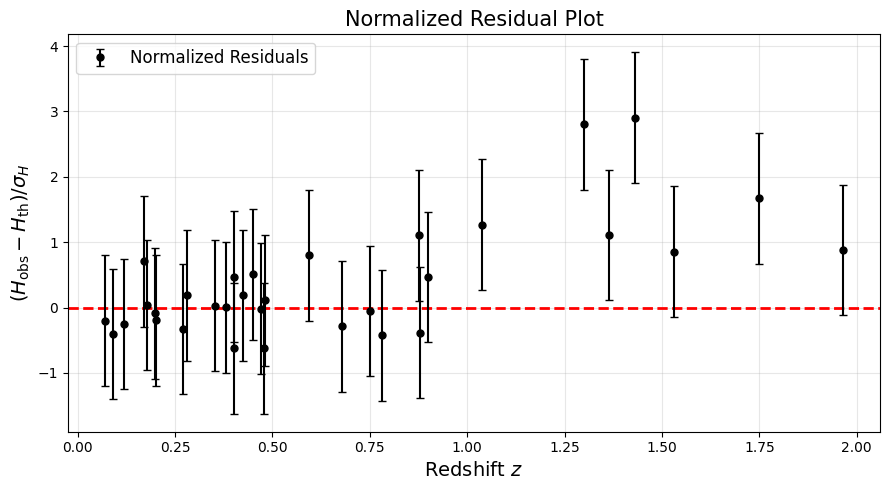}
    \caption{Normalized residuals of the CC fit.}
    \label{fig:Residual}
\end{subfigure}
\hfill
\begin{subfigure}[t]{0.48\textwidth}
    \centering
    \includegraphics[width=\textwidth]{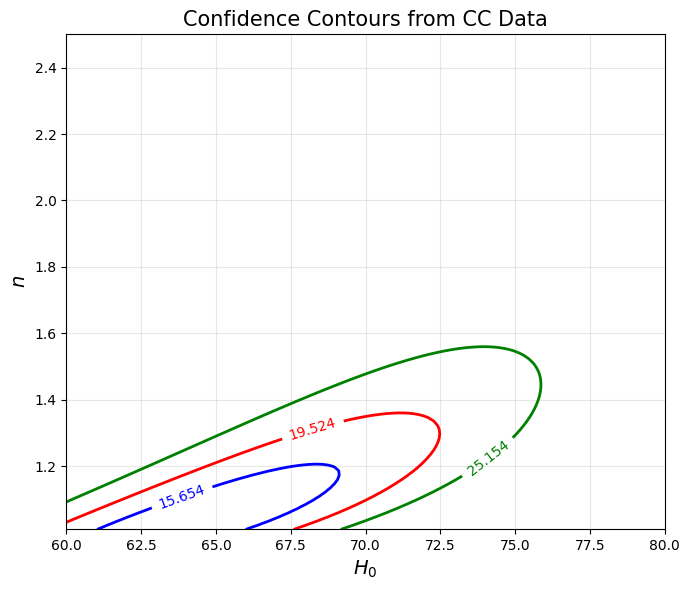}
    \caption{Confidence contours in the $(H_0,n)$ parameter space.}
    \label{fig:Contour}
\end{subfigure}

\caption{Observational constraints on the reconstructed power-law cosmological model using the 32 Cosmic Chronometer (CC) data. Panel (a) presents the comparison of the best-fit Hubble parameter with the observational measurements. Panel (b) shows the normalized residuals, indicating the goodness of fit between the theoretical model and the CC data. Panel (c) displays the corresponding $1\sigma$, $2\sigma$, and $3\sigma$ confidence contours in the $(H_0,n)$ parameter space obtained from the $\chi^2$ minimization.}
\label{fig:CC_analysis}
\end{figure}

Furthermore, for any physically viable modified gravity model within this framework, the viability conditions require the avoidance of ghost states and Laplacian instabilities, which translate to:
\begin{equation}
f_Q > 0,
\label{eq:cond1}
\end{equation}
and
\begin{equation}
f_{QQ} > 0.
\label{eq:cond2}
\end{equation}
For our polynomial model, these inequalities take the explicit forms:
\begin{equation}
1+\alpha mQ^{m-1}+\beta qQ^{q-1}> 0,
\label{eq:viability1}
\end{equation}
and
\begin{equation}
\alpha m(m-1)Q^{m-2}+\beta q(q-1)Q^{q-2}>0.
\label{eq:viability2}
\end{equation}
These bounds place strict constraints on the allowable parameter space $(\alpha, \beta, m, n)$ and preserve cosmic stability throughout the background evolution.

For the observational analysis, we use the 32 cosmic chronometer (CC) measurements of $H(z)$ compiled by Favale, G\'omez-Valent and Migliaccio~\cite{Favale2023}, covering the redshift range $0.07\leq z\leq1.965$. The reconstructed expansion history is confronted with 32 cosmic chronometer (CC) measurements over $0.07\leq z\leq1.965$ by minimising
\begin{equation}
\chi^2=\sum_{i=1}^{32}\frac{\left[H_{\rm obs}(z_i)-H_{\rm th}(z_i)\right]^2}{\sigma_i^2}.\nonumber
\end{equation}
As we are considering the power-law background leading to the Hubble parameter $H(z)=H_0(1+z)^{1/n}$, this yields the best-fit values $H_0=64.0295~\mathrm{km s^{-1}\,Mpc^{-1}}$ and $n=1.02782$. For the power-law background $a(t)=a_0t^n$ we find $H=n/t$, $\dot{H}=-n/t^2$ and $Q=6n^2/t^2$, while the pressureless matter sector is given by $p=0$ and $\rho=\rho_0a^{-3}$. Throughout the numerical analysis, we always use $\rho_0=\Omega_{m0}3H_0^2/(8\pi)$ with $\Omega_{m0}=0.315$. The resulting theoretical curve is compared with the CC observations in Fig.~\ref{fig:HzCC}. Fig.~\ref{fig:Residual} shows the normalized residuals, $(H_{\rm obs}-H_{\rm th})/\sigma$, for the 32 CC measurements. The residuals are widely scattered about zero without any obvious systematic trend with redshift, which indicates a good description of the observed expansion history. A few higher redshift points have relatively larger residuals but the overall distribution of the residuals do not show a significant systematic deviation of the reconstructed model from the CC data. In Fig.~\ref{fig:Contour} we have displayed the joint confidence contours on the $(H_0,n)$ parameter space. The three curves represent $1\sigma$, $2\sigma$, and $3\sigma$ confidence contours in the $(H_0,n)$ parameter space, that correspond to $\Delta\chi^2=2.30$, $6.17$, and $11.83$, respectively. The contours exhibit a positive correlation between $H_0$ and $n$, showing that the CC data permit a range of correlated values of these two parameters.This indicates a degeneracy between the two parameters in their joint determination from the CC data. The best-fit values $H_0=64.0295~\mathrm{km s^{-1}\,Mpc^{-1}}$ and $n=1.02782$ are in the allowed confidence region and confirm that the power-law background chosen satisfies CC constraints. We note that the CC analysis best fit value $H_0=64.0295~\mathrm{kms^{-1}\,Mpc^{-1}}$ is closer to the Planck $\Lambda$CDM determination, $H_0=67.4~\mathrm{kms^{-1}\,Mpc^{-1}}$, than to the local SH0ES determination, $H_0\simeq73~\mathrm{kms^{-1}\,Mpc^{-1}}$.

The evolution of the Dark Energy equation-of-state parameter $w_{DE}$ is determined from Eq.~(29) after substituting the power-law background $a(t)=a_0t^n$, the relations $H=n/t$ and $Q=6n^2/t^2$, the pressureless matter density $\rho=\rho_0(t/t_0)^{-3n}$ and the polynomial form $f(Q)=Q+\alpha Q^m+\beta Q^q$. The resulting time dependence is transformed to redshift by $t=t_0(1+z)^{-1/n}$ with $t_0=n/H_0$. The evolution is shown in Fig.~\ref{fig:wDE}.

\subsection*{Stability analysis}
The squared speed of sound $v_s^2=\dot{p}_{DE}/\dot{\rho}_{DE}$ is computed numerically by choosing the power-law scale factor $a(t)=a_0t^n$ which means $H=n/t$, $Q=6n^2/t^2$ and $\rho=\rho_0(t/t_0)^{-3n}$ and using Eqs.~\eqref{eq:N_def}-\eqref{eq:vs2}. For the polynomial form of $f(Q)=Q+\alpha Q^m+\beta Q^q$, the quantities $f_Q, f_{QQ}$ and $f_{QQQ}$ are computed analytically. As shown in Fig.~\ref{fig:vs2}, the squared speed of sound remains positive throughout the considered redshift range, indicating the classical stability of the model. The parameters $H_0=64.0295~\mathrm{kms^{-1}\,Mpc^{-1}}$ and $n=1.02782$ are fixed from the previous cosmic-chronometer analysis. The other parameters $(\alpha,\beta,m,q,\lambda)$ are chosen by a constrained numerical optimization under the viability conditions $f_Q>0$, $f_{QQ}>0$ and $0\leq v_s^2\leq1$ in the redshift range $0\leq z\leq1.965$. For the viable parameter set obtained, $v_s^2$ is positive in the entire redshift range considered, indicating the classical stability of the effective Dark Energy sector.

\subsection*{Note on the effective EoS}
It may be noted that $w_{\rm DE}$ considered here is the effective equation-of-state parameter related to the reconstructed modified-gravity sector. It is thus not necessarily expected to reproduce the value $w=-1$ corresponding to a minimally coupled cosmological constant in GR. In the allowed region of the parameter space that is constrained by the CC analysis and the stability conditions, $w_{\rm DE}$ is finite and slightly negative in the redshift interval considered. This behaviour should be considered in terms of the modified Friedmann dynamics rather than as an explicit determination of the cosmological-constant equation of state. The positive values of the corresponding squared sound speed also show that the reconstructed effective sector is classically stable for the same range of redshifts.

The theory presented in this Section is given by $f(Q,T)=f(Q)+\lambda T$. The corresponding results when $\lambda$ is kept are called the $f(Q,T)$ sector. However, in the limit of a vanishing matter-geometry coupling, $\lambda=0$, the theory reduces to the pure $f(Q)$ sector. While concluding the present section, we make an explicit distinction between these two cases in the following. The phenomenological parameterization of Eq.~\eqref{eq:polyfQ} is to be distinguished from the reconstruction scheme developed afterward. We use a polynomial form to get explicit expressions for the effective equation of state, the squared sound velocity and other cosmological diagnostics, which allows for a transparent analysis of the effects of nonlinear geometric corrections. In this sense, the parameters $\alpha$, $\beta$, $m$ and $q$ explored in this paper describe a generic nonlinear $f(Q)$ sector, and should not be regarded as the unique parameters for the reconstructed gravitational function of the GRGB-HDE.

\section{Generalized Ricci-Gauss-Bonnet Holographic Dark Energy with Nojiri-Odintsov Entropy} \label{sec:GRGBHDE}

The Dark Energy density is related to an infrared (IR) cutoff of an effective quantum field theory within the holographic principle \cite{Bousso2000,Bousso2002,Li2004}. In the present work, we consider a generalized Ricci-Gauss-Bonnet holographic Dark Energy (GRGB-HDE) scenario, in which the IR cutoff is constructed from the Ricci scalar and the Gauss-Bonnet invariant
\cite{Gao2009,Saridakis2018}. Following the Nojiri-Odintsov prescription \cite{Nojiri2022,NojiriOdintsovPaul2022}, we consider a modified entropy for the holographic horizon. The resulting entropy includes logarithmic corrections to the standard Bekenstein-Hawking area-entropy relation \cite{Bekenstein1973,Nojiri2022,NojiriOdintsovPaul2022}, and hence it provides us with a natural extension of the conventional holographic Dark Energy framework. 

\subsection{Ricci-Gauss-Bonnet Infrared Cutoff}

To incorporate both Ricci and Gauss-Bonnet contributions we choose the infrared scale as \begin{equation} L^{-2}=\alpha_1 R+\beta_1 G, \label{eq:RGBcutoff} \end{equation} where $\alpha_1$ and $\beta_1$ are constants that determine the relative contributions of Ricci and Gauss-Bonnet invariants \cite{Saridakis2018,Dubey2025}.
For a spatially flat FLRW spacetime, 
\begin{equation}
R=6(\dot H+2H^2), \qquad G=24H^2(\dot H+H^2).
\label{eq:RGBinvariants}
\end{equation}

Substituting Eq.~\eqref{eq:RGBcutoff} into the entropy-corrected holographic prescription \cite{Nojiri2022,NojiriOdintsovPaul2022}, we get the generalized Ricci-Gauss-Bonnet holographic Dark Energy density
\begin{equation} 
\rho_{\rm GRGB}=3c^2M_p^2\left(\alpha_1R+\beta_1G\right)\left[1+\eta\ln\left(\frac{4\pi}{A_0(\alpha_1R+\beta_1G)}\right)\right]. 
\label{eq:GRGBdensity} 
\end{equation}

Equation~\eqref{eq:GRGBdensity} represents the Dark Energy sector which will be reconstructed afterward in the $f(Q,T)$ framework. It is useful to note that the generalized construction contains several limiting cases. In particular, setting $\beta_1=0$ we obtain the Ricci holographic model and suppressing the logarithmic correction $\eta=0$ we obtain the corresponding uncorrected Ricci-Gauss-Bonnet holographic density.

\subsection{Correspondence with the Effective \texorpdfstring{$f(Q,T)$}{f(Q,T)} Sector}

The effective Dark Energy density obtained from the modified Friedmann equations of $f(Q,T)$ theory is set equal to the GRGB-HDE density \cite{NojiriOdintsov2017,NojiriOdintsovPaul2022,Dubey2025},
\begin{equation}
\rho_{\rm DE}=\rho_{\rm GRGB}.
\label{eq:correspondence}
\end{equation}
This correspondence is the basis for reconstructing the gravitational function consistent with the given holographic Dark Energy density.

It is important to separate the two ingredients that go into this construction. The quantity $\rho_{\rm GRGB}$ is the holographic Dark Energy prescription thru the curvature-based infrared cutoff and the Nojiri-Odintsov entropy correction \cite{Saridakis2018,Dubey2025,Nojiri2022,NojiriOdintsovPaul2022}, while $\rho_{\rm DE}$ is the effective cosmological sector generated by the $f(Q,T)$ gravitational dynamics. Their correspondence thus allows to translate the holographic description into a reconstructed modified-gravity model.

\subsection{\texorpdfstring{Reconstruction of $f(Q,T)$ Gravity}{Reconstruction of f(Q,T) Gravity}}
\label{subsec:reconstruction}

In the present section, instead of assuming any functional form of $f(Q)$, we implement a reconstruction method, where the GRGB-HDE density is equated to the effective Dark Energy density obtained from the modified Friedmann equations as the reconstruction approach. This correspondence produces a first-order differential equation for $f(Q)$ which we solve numerically. The trace contribution is then included to the resulting geometric function to obtain the full reconstructed $f(Q,T)$ theory. Thus, the polynomial parametrization discussed above and the GRGB-HDE reconstruction are two complementary levels of description: the first one offers an explicit phenomenological description of the nonlinear geometric corrections, whereas the second one is used to determine the gravitational function from the holographic correspondence without setting a prior functional form.The correspondence in Eq.~\eqref{eq:correspondence} provides the basis for reconstructing the gravitational function associated with the generalized Ricci-Gauss-Bonnet holographic dark energy prescription. In the present framework, the gravitational function
is written as
\begin{equation}
f(Q,T)=f(Q)+\lambda T,
\label{Eq:fQ_T}
\end{equation}
where the first term represents the purely geometric
non-metricity sector, while the second term accounts for the
non-minimal matter-geometry coupling.

Using the effective Dark Energy density obtained from the modified
Friedmann equations, Eq.~\eqref{eq:rhoDE}, the correspondence
condition can be written as
\begin{equation}
\rho_{\rm GRGB}=\frac{1}{32\pi f_Q}\left[f-16\pi\rho-2\lambda(\rho+p)-32\pi f_Q\rho\right].
\label{Eq:Correspondence}
\end{equation}
For the pressureless matter configuration considered in this work,
$p=0$, and hence
\begin{equation}
\rho_{\rm GRGB}=\frac{1}{32\pi f_Q}\left[f-16\pi\rho-2\lambda\rho-32\pi f_Q\rho\right].
\end{equation}
Rearranging the above equation gives
\begin{equation}
f-32\pi\left(\rho+\rho_{\rm GRGB}\right)f_Q -(2\lambda+16\pi)\rho=0.
\label{eq:reconstruction_ode}
\end{equation}
Equation~\eqref{eq:reconstruction_ode} represents the reconstruction equation for the purely geometric function $f(Q)$ in the presence of the matter-geometry coupling. Thus, although the reconstructed function depends on the non-metricity scalar $Q$, the reconstruction is performed within the full $f(Q,T)$ framework through the explicit $\lambda$-dependent contribution.

For the power-law background
\begin{equation}
a(t)=a_0t^n,\qquad H=\frac{n}{t}, \qquad Q=\frac{6n^2}{t^2},
\label{Eq:powerlaw}
\end{equation}
the cosmic time can be expressed in terms of the non-metricity scalar as
\begin{equation}
t=\frac{\sqrt{6}\,n}{\sqrt{Q}}.
\end{equation}
Consequently, the matter density
$\rho=\rho_0a^{-3}$ and the curvature invariants entering the GRGB-HDE density can be expressed as functions of $Q$. In
particular,
\begin{equation}
R=6(\dot H+2H^2)
=\frac{6n(2n-1)}{t^2},
\end{equation}
and
\begin{equation}
G=24H^2(\dot H+H^2)
=\frac{24n^3(n-1)}{t^4}.
\end{equation}
Using $Q=6n^2/t^2$, these quantities take the form
\begin{equation}
R(Q)=\frac{2n-1}{n}\,Q,
\label{Eq:R}
\end{equation}
and
\begin{equation}
G=\frac{2(n-1)}{3n}\,Q^2.
\label{Eq:G}
\end{equation}
Hence, the generalized Ricci-Gauss-Bonnet holographic density $\rho_{\rm GRGB}$ can be expressed entirely in terms of $Q$ for the adopted power-law background.

The reconstruction equation~\eqref{eq:reconstruction_ode} can therefore be treated as a first-order differential equation for
$f(Q)$. Once the holographic density and the matter density are expressed in terms of $Q$, the resulting equation is solved subject to the appropriate boundary condition. The complete gravitational function is subsequently recovered from $f(Q,T)$ as in Eq.~\eqref{Eq:fQ_T}.

It is important to emphasize that the reconstruction separates the two physical contributions. The function $f(Q)$ describes the geometric modification associated with the non-metricity scalar, whereas $\lambda T$ represents the explicit matter-geometry coupling. The limit $\lambda=0$ consequently reduces the reconstructed theory to the corresponding pure $f(Q)$ model.

At this stage, to obtain an explicit differential equation for the gravitational function $f(Q)$, we first express the matter density and the GRGB-HDE density in terms of the non-metricity scalar $Q$. For the pressureless matter sector, we have
$\rho=\rho_0 a^{-3}$. Using the power-law scale factor $a(t)=a_0t^n$ together with $Q=\frac{6n^2}{t^2}$, we can have 
$t=\frac{\sqrt{6}\,n}{\sqrt{Q}}$, and consequently
\begin{equation}
\rho(Q) = \rho_0a_0^{-3} \left(\frac{Q}{6n^2}\right)^{3n/2}.
\label{eq:rhoQ_reconstruction}
\end{equation}
Similarly, from Eqs.~\eqref{Eq:R} and \eqref{Eq:G}, the Ricci scalar and the Gauss-Bonnet invariant can be written as
\begin{equation}
R(Q)=\frac{2n-1}{n}\,Q,
\qquad
G(Q)=\frac{2(n-1)}{3n}\,Q^2.
\label{eq:RG_Q}
\end{equation}
It is therefore convenient to define
\begin{equation}
{\cal X}(Q) \equiv \alpha_1R(Q)+\beta_1G(Q),
\end{equation}
which gives
\begin{equation}
{\cal X}(Q) = \frac{\alpha_1(2n-1)}{n}\,Q + \frac{2\beta_1(n-1)}{3n}\,Q^2.
\label{eq:XQ}
\end{equation}
Using Eq.~\eqref{eq:GRGBdensity}, the generalized Ricci-Gauss-Bonnet holographic Dark Energy density consequently becomes
\begin{equation}
\rho_{\rm GRGB}(Q) = 3c^2M_p^2{\cal X}(Q) \left[1+\eta\ln \left(\frac{4\pi}{A_0{\cal X}(Q)}\right)\right].
\label{eq:rhoGRGBQ}
\end{equation}
After substitution of Eqs.~\eqref{eq:rhoQ_reconstruction} and \eqref{eq:rhoGRGBQ} into the reconstruction equation \eqref{eq:reconstruction_ode}, with $f_Q=df/dQ$, we get the following first-order differential equation:
\begin{equation}
\begin{aligned}
f(Q)&-32\pi \Bigg[\rho_0a_0^{-3} \left(\frac{Q}{6n^2}\right)^{3n/2} \\
&\qquad +3c^2M_p^2{\cal X}(Q) \left\{1+\eta\ln\left[\frac{4\pi}{A_0X(Q)} \right]\right\}\Bigg]
\frac{df(Q)}{dQ}\\
&-(2\lambda+16\pi) \rho_0a_0^{-3} \left(\frac{Q}{6n^2}\right)^{3n/2}=0,
\end{aligned}
\label{eq:explicit_reconstruction}
\end{equation}
where ${\cal X}(Q)$ is given by Eq.~\eqref{eq:XQ}.

Equivalently, Eq.~\eqref{eq:explicit_reconstruction} can be written
in the standard first-order linear form
\begin{equation}
\frac{df(Q)}{dQ} -\frac{f(Q)}{32\pi\left[\rho(Q)+\rho_{\rm GRGB}(Q)\right]}=-\frac{(8\pi+\lambda)\rho(Q)}
{16\pi\left[\rho(Q)+\rho_{\rm GRGB}(Q)\right]}.
\label{eq:linear_reconstruction}
\end{equation}
Equation~\eqref{eq:linear_reconstruction} constitutes the reconstruction equation for the geometric function $f(Q)$ in the
present $f(Q,T)$ framework. Equation~\eqref{eq:linear_reconstruction} is a first-order linear differential equation for the geometric function $f(Q)$. For convenience, we introduce
\begin{equation}
\mathcal{D}(Q) \equiv \rho(Q) + \rho_{\text{GRGB}}(Q),
\end{equation}
so that Eq.~\eqref{eq:linear_reconstruction} takes the form
\begin{equation}
\frac{\mathrm{d}f}{\mathrm{d}Q} - \frac{1}{32\pi\mathcal{D}(Q)}f = -\frac{(8\pi+\lambda)\rho(Q)}{16\pi\mathcal{D}(Q)}.
\label{Eq:ODE}
\end{equation}

To obtain the integrating factor in a form suitable for the
reconstruction, we introduce a reference value $Q_*$ and define
\begin{equation}
I(Q)\equiv
\int_{Q_*}^{Q}
\frac{\mathrm{d}\widetilde{Q}}
{32\pi\mathcal{D}(\widetilde{Q})}.
\label{62}
\end{equation}
The integrating factor is therefore
\begin{equation}
\mu(Q)=e^{-I(Q)}.
\label{63}
\end{equation}
By construction, $\mu(Q_*)=1$. Multiplying Eq.~\eqref{Eq:ODE} by this
integrating factor gives
\begin{equation}
\frac{\mathrm{d}}{\mathrm{d}Q}
\left[\mu(Q)f(Q)\right] =- \mu(Q) \frac{(8\pi+\lambda)\rho(Q)} {16\pi\mathcal{D}(Q)}.
\label{64}
\end{equation}
Integrating from $Q_*$ to $Q$, and denoting
$f(Q_*)\equiv f_*$, yields
\begin{equation}
f(Q)=e^{I(Q)} \left[f_*- \frac{8\pi+\lambda}{16\pi} \int_{Q_*}^{Q} \frac{\rho(\widetilde{Q})} {\mathcal{D}(\widetilde{Q})}
e^{-I(\widetilde{Q})} \,\mathrm{d}\widetilde{Q}\right].
\label{Eq:fQ}
\end{equation}
For the full GRGB-HDE model considered here, the logarithmic Nojiri-Odintsov correction prevents the integral defining $I(Q)$
from, in general, admitting an elementary closed-form antiderivative. We therefore evaluate $I(Q)$ and the remaining
integral in Eq.~\eqref{Eq:fQ} numerically. The boundary value $f_*$ is fixed by the boundary condition adopted for the reconstructed gravitational function. Using the definition of the trace in Eq.~\eqref{eq:Ttrace},
\begin{equation}
T = -\rho + 3p.
\end{equation}
For the pressureless matter sector considered throughout the reconstruction, $p = 0$, and hence \eqref{eq:Ttrace} reduces to
\begin{equation}
T = -\rho.
\end{equation}
Using Eq.~\eqref{eq:rhoQ_reconstruction}, we consequently obtain
\begin{equation}
T(Q) = -\rho_0 a_0^{-3} \left(\frac{Q}{6n^2}\right)^{3n/2}.
\label{eq:TQ_final}
\end{equation}

Combining this result with the reconstructed geometric function given by Eq.~\eqref{Eq:fQ}, the complete reconstructed gravitational function takes the form
\begin{equation}
\begin{aligned}
f(Q,T) ={} & e^{I(Q)} \left[ f_* - \frac{8\pi+\lambda}{16\pi} \int_{Q_*}^{Q} \frac{\rho(\tilde{Q})}{D(\tilde{Q})} e^{-I(\tilde{Q})} \,\mathrm{d}\tilde{Q} \right] \\
& -\lambda \rho_0 a_0^{-3} \left(\frac{Q}{6n^2}\right)^{3n/2},
\end{aligned}
\label{eq:full_fQT}
\end{equation}
where
\begin{equation}
I(Q) = \int_{Q_*}^{Q} \frac{\mathrm{d}\tilde{Q}}{32\pi D(\tilde{Q})}, \qquad D(Q) = \rho(Q) + \rho_{\text{GRGB}}(Q).
\label{Eq:IQ}
\end{equation}

Equation~\eqref{eq:full_fQT} represents the reconstructed $f(Q,T)$ gravitational function, with the first term corresponding to the reconstructed geometric sector and the second term arising explicitly from the matter-geometry coupling $\lambda T$ presented as a function of $Q$.

\begin{figure*}[t]
    \centering

    \begin{subfigure}[b]{0.32\textwidth}
        \centering
        \includegraphics[width=\textwidth]{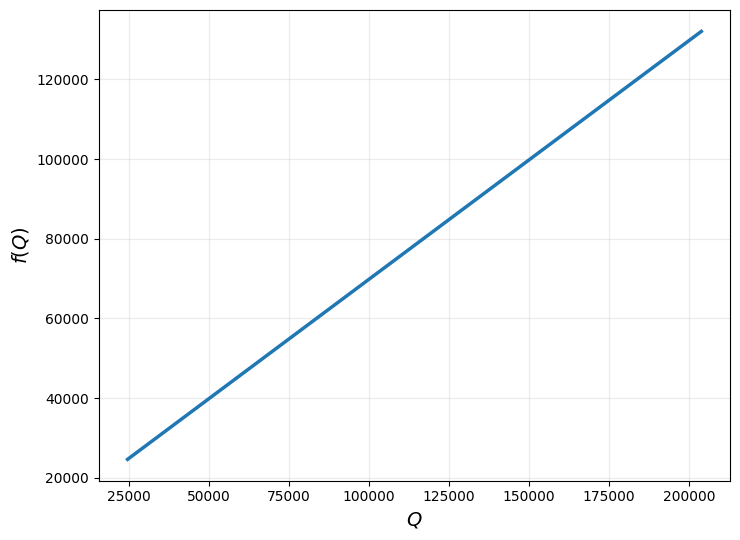}
        \caption{}
        \label{fig:fQ}
    \end{subfigure}
    \hfill
    \begin{subfigure}[b]{0.32\textwidth}
        \centering
        \includegraphics[width=\textwidth]{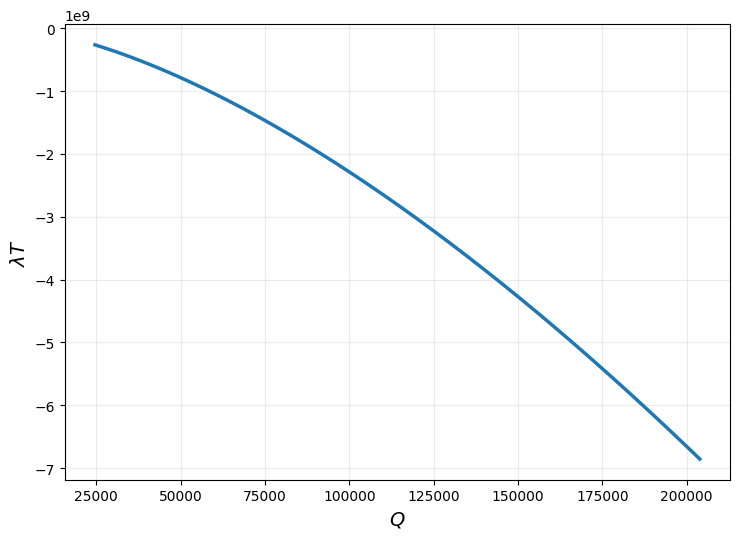}
        \caption{}
        \label{fig:lambdaT}
    \end{subfigure}
    \hfill
    \begin{subfigure}[b]{0.32\textwidth}
        \centering
        \includegraphics[width=\textwidth]{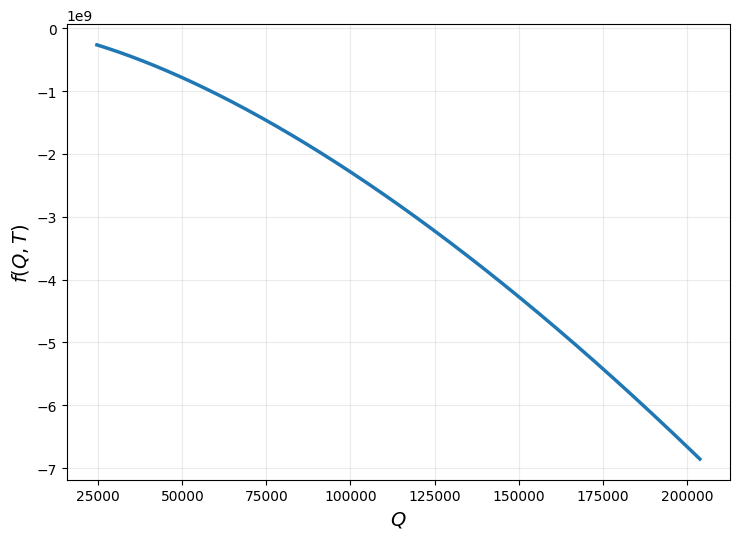}
        \caption{}
        \label{fig:fQT}
    \end{subfigure}

    \caption{Reconstructed gravitational functions for $f(Q,T)$ model with CC-constraint. The left, middle and right subfigures show, respectively, the reconstructed geometric function $f(Q)$, the matter-trace contribution $\lambda T$ and the complete function $f(Q,T)=f(Q)+\lambda T$ as functions of the non-metricity scalar $Q$. The reconstruction is based on $H_0=64.0295~\mathrm{kms^{-1}\,Mpc^{-1}}$, $n=1.02782$, $\lambda=4.9986$, $\alpha_1=8.38112$, $\beta_1=0.00408079$, $c=0.0443364$, $\eta=1.19592$ and $A_0=4.54691\times10^{-6}$.}
    \label{fig:fQT_diagnostics}
\end{figure*}

For the observational reconstruction the background expansion was firstly constrained using the 32 Cosmic Chronometer (CC) measurements \cite{Favale2023} for which the power-law form $H(z)=H_0(1+z)^{1/n}$ gives $H_0=64.0295~\mathrm{kms^{-1}\,Mpc^{-1}}$ and $n=1.02782$. The GRGB-HDE parameters entering $\rho_{\rm GRGB}$ were then optimized by the CC-constrained Dark Energy density with fixed background parameters. The resulting set of parameters is $\alpha_1=8.38112$, $\beta_1=0.00408079$, $c=0.0443364$, $\eta=1.19592$, and $A_0=4.54691\times10^{-6}$. The matter-geometry coupling is fixed at $\lambda=4.9986$. The background expansion is first constrained by the 32 CC measurements,
yielding $H_0=64.0295~\mathrm{kms^{-1}\,Mpc^{-1}}$ and $n=1.02782$, with $\chi^2_{\rm CC}=13.35$ for 30 degrees of freedom
and $\chi^2_{\rm CC,red}=0.445$. Keeping these background parameters fixed, the GRGB-HDE parameters are subsequently determined by matching the corresponding effective Dark Energy density to the reconstructed background density. This latter step is used as a numerical reconstruction procedure rather than as an independent observational $\chi^2$ constraint. The optimized parameters, together with the CC-constrained values of $H_0$ and $n$, are then used to evaluate $Q(z)=6H^2(z)$ and to reconstruct the $ f(Q,T)$ function from Eqs~\eqref{eq:full_fQT}-\eqref{Eq:IQ} numerically without imposing any other functional ansatz for $f(Q)$. It may be noted that the reconstruction scheme involves the parameters $\alpha_1$ and $\beta_1$ through the geometric linear combination $X(Q) = \alpha_1 R(Q) + \beta_1 G(Q),$ where $R(Q)$ and $G(Q)$ are given in Eq. \eqref{eq:RG_Q}. The parameters $c$, $\eta$ and $A_0$ are defined via the GRGB holographic Dark Energy density, which is included in the total density function $D(Q) = \rho(Q) + \rho_{\mathrm{GRGB}}(Q)$. This means that $\alpha_1$, $\beta_1$, $c$, $\eta$, and $A_0$ are not present in the final reconstructed solution, Eq.~Eqs~\eqref{eq:full_fQT}, explicitly, but they influence it implicitly through $D(Q)$. In this way the integrating factor $I(Q)$ is computed and the function $f(Q,T)$ gets numerical values for plotting.

Let us now look at Fig. \ref{fig:fQT_diagnostics}. The negative slope of the reconstructed $f(Q,T)$, as seen in the right most panel, with respect to $Q$ can not be interpreted as a violation of the usual $f_Q>0$ viability condition. In this context, if we look at the reconstructed function is related to the geometric contribution through $f(Q,T)=f(Q)+\lambda T$, we can easily understand that for pressureless matter, $T=-\rho(Q)$, and hence along the cosmological background one has $f(Q,T(Q))=f(Q)-\lambda\rho(Q)$. Consequently, we can clearly have the cosmological trajectory  as $\frac{d f(Q,T(Q))}{dQ} = \frac{df(Q)}{dQ} -\lambda\frac{d\rho(Q)}{dQ}$. Since $\rho(Q)\propto Q^{3n/2}$, it follows that $d\rho/dQ>0$. Moreover, as for the present reconstruction we get the constrained value $\lambda=4.9986>0$, the trace-coupling contribution $-\lambda\,d\rho/dQ <0 $. It can therefore dominate over the positive geometric contribution $df/dQ$, resulting in a decreasing $f(Q,T)$ as observed in the first panel of Fig. \ref{fig:fQT_diagnostics} even when $f_Q>0$. Thus, the negative slope of the complete $f(Q,T)$ curve does not, by itself, imply a violation of the $f_Q>0$ viability condition. The reconstructed functions shown in Fig.~\ref{fig:fQT_diagnostics} provide a useful representation of the separate contributions to the modified gravitational sector. The geometrical contribution is implied by $f(Q)$ and it shows a monotone increasing behaviour with the non-metricity scalar $Q$. On the other hand, the trace-related contribution $\lambda T$ stays at negative level and becomes more prominent with increasing $Q$. Their combination leads the reconstructed $f(Q,T)$, to exhibit a decreasing trend over the considered range. This behaviour reflects the relative contributions of the geometrical and matter-trace sectors within the reconstructed model. In particular, from the monotone decreasing behaviour of the $f(Q,T)$ is not directly associated with the sign of $f_Q$, since $f(Q,T)$ contains the additional $Q$-dependent trace contribution through $T=-\rho(Q)$.

\subsection{Relation to generalized entropic cosmology}

Recently, the correspondence between generalized entropic cosmology and modified gravity based on non-metricity has been established. In particular, Nojiri and Odintsov \cite{NojiriOdintsov2025FQ} proved that the modified Friedmann equations derived from generalized horizon entropies can be recovered by a suitable $F(Q)$ gravity theory, setting up a correspondence between generalized entropic cosmology and $F(Q)$ gravity.This correspondence has been recently further investigated within the framework of inflation, where the generalized entropic cosmology and the $f(Q)$ gravity give equivalent descriptions for the construction of inflationary scenarios \cite{NojiriOdintsov2026FQ}.

The above correspondence offers an interesting interpretation of the reconstruction carried out in the present work. Our framework can also be interpreted within the framework of generalized entropic cosmology since the geometric sector of our theory is described by $f(Q)$ while the holographic sector is constructed by employing a generalized entropy prescription. The additional trace contribution $\lambda T$ in $f(Q,T)=f(Q)+\lambda T$, however, introduces a matter-geometry coupling that is not present in the pure $f(Q)$ correspondence. Thus, the present $f(Q,T)$ framework can be regarded as an extension of the generalized entropic-$f(Q)$ correspondence to a matter-geometry coupled sector.

\section{Thermodynamic behaviour with Nojiri-Odintsov entropy} \label{sec:Thermodynamics}
The thermodynamic description of an FLRW universe can be naturally formulated at the apparent horizon. In this framework, the Friedmann equations can be related to the first law of thermodynamics, thereby establishing a useful connection between gravitational dynamics and horizon thermodynamics \cite{CaiKim2005}. This approach provides a convenient framework for the analysis of the thermodynamic behavior of cosmological models with modified gravitational dynamics. Recent studies have further investigated the second law of horizon thermodynamics during cosmic evolution and the consistency of horizon
entropy with the FLRW equations in general modified theories of gravity \cite{OdintsovPaulSenGupta2024,NojiriOdintsovPaulSenGupta2024}. This approach provides a convenient framework for the analysis of the thermodynamic behaviour of cosmological models with modified gravitational dynamics.The thermodynamic description of the Friedmann equations at the apparent horizon was further investigated by including the horizon temperature and entropy as well as the effective energy content of the universe \cite{AkbarCai2007}. Such a formulation is particularly useful in modified gravity, where the gravitational sector can be described by effective energy density and pressure. In the present work we study this effective-fluid perspective in the investigation of the thermodynamic behavior of the reconstructed $f(Q,T)$ model. The generalized second law of thermodynamics gives an extra consistency condition for cosmological models. In particular, the total entropy of the horizon and the cosmic fluid can be studied to check whether the total entropy is a non-decreasing function during the cosmic evolution \cite{IzquierdoPavon2006} . In line with this thermodynamic perspective, we investigate the evolution of the total entropy in the reconstructed $f(Q,T)$ model and examine the validity of the generalized second law of thermodynamics. The thermodynamic description of an FLRW universe can be formulated by associating the apparent horizon with a thermodynamic boundary, thereby establishing a connection between horizon
thermodynamics and the Friedmann equations \cite{CaiKim2005,AkbarCai2007}.
Having reconstructed the gravitational function $f(Q,T)$ using the GRGB-HDE density incorporating the Nojiri-Odintsov logarithmic correction, we now examine the thermodynamic behaviour of the corresponding cosmological background. We treat the apparent horizon as the thermodynamic boundary, with its radius and area defined as
\begin{equation}
R_{\mathrm{A}} = \frac{1}{H}, \qquad
A = 4\pi R_{\mathrm{A}}^{2} = \frac{4\pi}{H^{2}}.
\end{equation}
The entropy correction approach adopted in the current study is motivated by the generalized entropy framework of Nojiri, Odintsov and Faraoni \cite{NojiriOdintsovFaraoni2022}, which demonstrates that replacing the standard Bekenstein-Hawking area law can modify the holographic Dark Energy sector and, consequently, the cosmological dynamics \cite{NojiriOdintsovFaraoni2022}. Recent developments in
entropic cosmology have further extended this framework to different aspects of cosmic evolution and to the holographic description from inflation through reheating \cite{NojiriOdintsovPaul2024,OdintsovDOnofrioPaul2023}. Following this entropy-based approach, we consider a logarithmically corrected horizon entropy of the form
\begin{equation}
S_{\mathrm{NO}} =\frac{\pi}{H^{2}}\left[1+\eta\ln\left(\frac{4\pi}{A_{0}H^{2}}\right)\right].
\label{eq:SNO}
\end{equation}
Here, the logarithmic correction strength is controlled by the parameter
$\eta$, whereas $A_{0}$ sets the reference area scale. The generalized second law is subsequently examined by considering the evolution of the total entropy of the horizon and the matter/energy content enclosed within it \cite{IzquierdoPavon2006}.

For the adopted power-law background as elaborated in Eq.~\eqref{Eq:powerlaw}, we have $a(t) = a_{0}t^{n},~H = \frac{n}{t}, ~\dot{H} = -\frac{H^{2}}{n}$ and the Hubble parameter evolves with redshift as $H(z) = H_{0}(1+z)^{1/n}$. Differentiating $S_{\mathrm{NO}}$ with respect to cosmic time yields
\begin{equation}
\dot{S}_{\mathrm{NO}} = -\frac{2\pi\dot{H}}{H^{3}} \left[ 1 + \eta \left\{ 1 + \ln \left( \frac{4\pi}{A_{0}H^{2}} \right) \right\} \right],
\end{equation}
which, under power-law evolution, simplifies to
\begin{equation}
\dot{S}_{\mathrm{NO}} = \frac{2\pi}{nH} \left[ 1 + \eta \left\{ 1 + \ln \left( \frac{4\pi}{A_{0}H^{2}} \right) \right\} \right].
\label{eq:SdotNO}
\end{equation}

It should be mentioned that the present thermodynamic analysis is motivated by the generalized entropy framework developed by Nojiri, Odintsov and Faraoni \cite{Nojiri2022}, in which departures from the Bekenstein-Hawking area law are considered and alternative entropy prescriptions are incorporated into holographic cosmology \cite{Nojiri2022}. In particular, it is demonstrated in their analysis that modifications of the horizon entropy can induce corresponding modifications in the holographic Dark Energy sector and thereby affect the cosmological dynamics. Inspired by this entropy-based
view of holographic cosmology \cite{Nojiri2022}, we consider a logarithmically corrected horizon entropy in the present work and investigate its consequences within the reconstructed $f(Q,T)$ framework.

To test the generalized second law of thermodynamics (GSLT), we incorporate the entropy of the effective cosmic fluid. From the effective Friedmann equations [Eqs.~\eqref{eq:einstein1} and~\eqref{eq:einstein2}], the total density and pressure are given by
\begin{equation}
\rho_{\mathrm{tot}} = \rho + \rho_{\mathrm{DE}} = \frac{3H^{2}}{8\pi},
\end{equation}
and
\begin{equation}
p_{\mathrm{tot}} = p + p_{\mathrm{DE}} = -\frac{1}{8\pi} \left( 2\dot{H} + 3H^{2} \right).
\end{equation}
The apparent horizon temperature and its enclosed volume are defined respectively as
\begin{equation}
T_{\mathrm{A}} = \frac{1}{2\pi R_{\mathrm{A}}} = \frac{H}{2\pi}, \qquad V = \frac{4\pi}{3} R_{\mathrm{A}}^{3} = \frac{4\pi}{3H^{3}}.
\end{equation}

The entropy evolution of the effective fluid follows from the Gibbs relation,
\begin{equation}
T_{\mathrm{A}} \dot{S}_{\mathrm{m}} = V \dot{\rho}_{\mathrm{tot}} + \left( \rho_{\mathrm{tot}} + p_{\mathrm{tot}} \right) \dot{V}.
\label{Eq:gibbs}
\end{equation}
Substituting the expressions for $\rho_{\mathrm{tot}}$ and $p_{\mathrm{tot}}$ leads to
\begin{equation}
\dot{S}_{\mathrm{m}} = \frac{2\pi\dot{H}}{H^{5}} \left( H^{2} + \dot{H} \right),
\end{equation}
which, for the power-law background, reduces to
\begin{equation}
\dot{S}_{\mathrm{m}} = -\frac{2\pi(n-1)}{n^{2}H}.
\end{equation}

The total entropy of the system is $S_{\mathrm{tot}} = S_{\mathrm{NO}} + S_{\mathrm{m}}$, and the GSLT mandates that
\begin{equation}
\dot{S}_{\mathrm{tot}} = \dot{S}_{\mathrm{NO}} + \dot{S}_{\mathrm{m}} \ge 0.
\end{equation}
Combining the individual contributions gives the total rate of entropy change:
\begin{equation}
\dot{S}_{\mathrm{tot}} = \frac{2\pi}{nH} \left[ 1 + \eta \left\{ 1 + \ln \left( \frac{4\pi}{A_{0}H^{2}} \right) \right\} 
\right] - \frac{2\pi(n-1)}{n^{2}H}.
\label{Eq:81}
\end{equation}
By expressing $H$ in terms of redshift via $H(z) = H_{0}(1+z)^{1/n}$, the entropy evolution can be numerically evaluated over the redshift domain used in the observational and stability analyses.

For the CC-constrained background, we adopt
\begin{equation}
H_{0} = 64.0295\,\mathrm{kms^{-1}\,Mpc^{-1}}, \qquad n = 1.02782,
\end{equation}
along with the reconstructed Nojiri-Odintsov parameters
\begin{equation}
\eta = 1.19592, \qquad A_{0} = 4.54691 \times 10^{-6}.
\end{equation}
These parameters mentioned above are directly taken from the GRGB-HDE reconstruction and are not independently fitted again for the thermodynamics here. The behaviour of $S_{\mathrm{NO}}$ and $\dot{S}_{\mathrm{tot}}$ are therefore considered to provide a distinct test for the thermodynamic consistency of the reconstructed model elaborated in the previous section. Now we shall test the validation limit for the GSLT i.e. $\dot{S}_{\mathrm{tot}} \ge 0$ over the range of redshifts considered.

\begin{figure*}[t]
    \centering
    \begin{subfigure}[b]{0.32\textwidth}
        \centering
        \includegraphics[width=\textwidth]{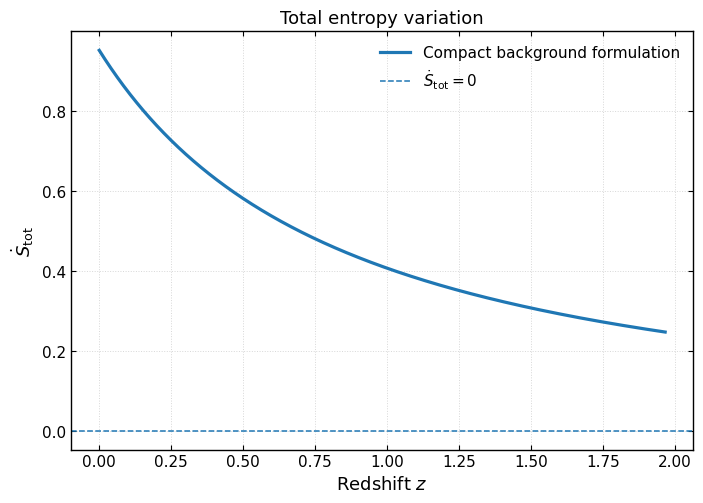}
        \caption{}
        \label{fig:entropy_compact}
    \end{subfigure}
    \hfill
    \begin{subfigure}[b]{0.32\textwidth}
        \centering
        \includegraphics[width=\textwidth]{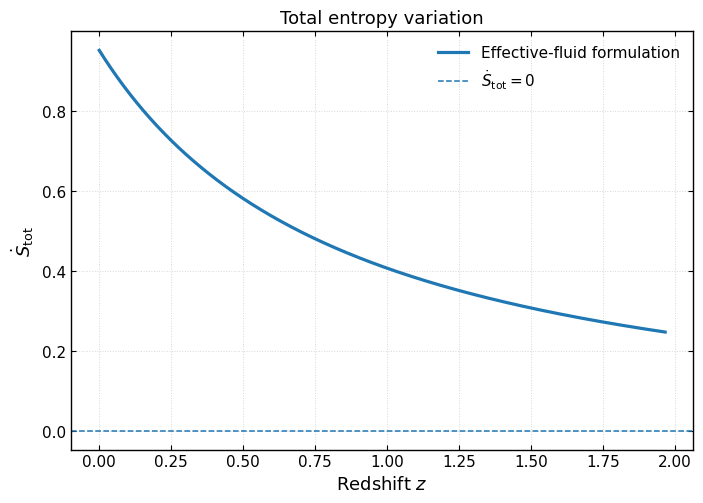}
        \caption{}
        \label{fig:entropy_fqt}
    \end{subfigure}
    \hfill
    \begin{subfigure}[b]{0.32\textwidth}
        \centering
        \includegraphics[width=\textwidth]{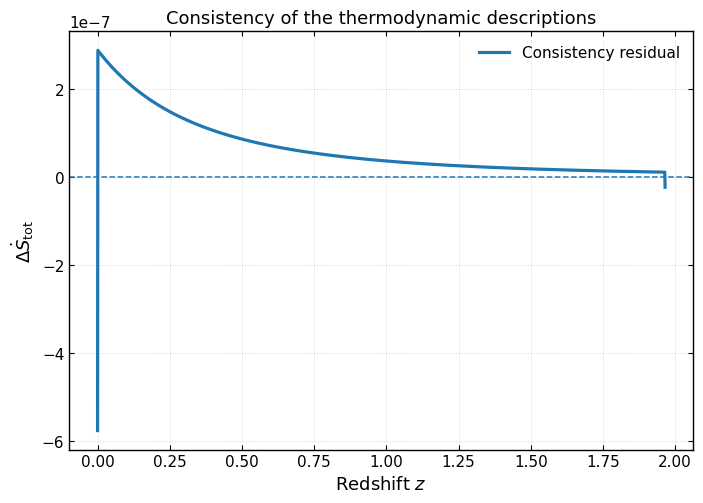}
        \caption{}
        \label{fig:entropy_residual}
    \end{subfigure}

    \caption{
    Evolution of the total entropy variation and its consistency within the reconstructed $f(Q,T)$ cosmology. Panel (a) shows
    $\dot{S}_{\rm tot}$ obtained from the compact background description, while panel (b) presents the corresponding result from the effective-fluid formulation associated with the reconstructed $f(Q,T)$ framework. Panel (c) shows the consistency
    residual $\Delta\dot{S}_{\rm tot}=\dot{S}_{\rm tot}^{(f(Q,T))}-\dot{S}_{\rm tot}^{\rm bg}$.
    The horizontal dashed line in panels (a) and (b) denotes $\dot{S}_{\rm tot}=0$.}
    \label{fig:entropy_fQT_consistency}
\end{figure*}

At this stage, in order to establish a connection between the thermodynamic analysis and the reconstructed $f(Q,T)$ gravity as already obtained, we identify the total effective cosmic fluid as
\begin{equation}
\rho_{\mathrm{tot}} = \rho + \rho_{\mathrm{DE}}^{(f(Q,T))}, \qquad p_{\mathrm{tot}} = p + p_{\mathrm{DE}}^{(f(Q,T))},
\label{Eq:84}
\end{equation}
where $\rho_{\mathrm{DE}}^{(f(Q,T))}$ and $p_{\mathrm{DE}}^{(f(Q,T))}$ represent the effective Dark Energy density and pressure as already defined in Eqs.~\eqref{eq:rhoDE} and~\eqref{eq:pDE}, respectively. We have written the quantities by rewriting the modified Friedmann equations in an Einstein-like form. This takes into account both the effects of modified geometry and the nonminimal matter-geometry coupling. Applying the Gibbs relation with the apparent horizon as the enveloping horizon, the time-derivative of the entropy for the effective cosmic fluid, as written in \eqref{Eq:gibbs} is now re-written as
\begin{equation}
T_{\mathrm{A}} \dot{S}_{\mathrm{m}} = V \frac{\mathrm{d}}{\mathrm{d}t} \left[ \rho + \rho_{\mathrm{DE}}^{(f(Q,T))} \right] + \left[ \rho + p + \rho_{\mathrm{DE}}^{(f(Q,T))} + p_{\mathrm{DE}}^{(f(Q,T))} \right] \dot{V}.
\label{Eq:85}
\end{equation}
Thus, we have the total entropy variation in the following form :
\begin{equation}
\dot{S}_{\mathrm{tot}} = \dot{S}_{\mathrm{NO}} + \frac{1}{T_{\mathrm{A}}} \left\{ V \frac{\mathrm{d}}{\mathrm{d}t} \left[ \rho + \rho_{\mathrm{DE}}^{(f(Q,T))} \right] + \left[ \rho + p + \rho_{\mathrm{DE}}^{(f(Q,T))} + p_{\mathrm{DE}}^{(f(Q,T))} \right] \dot{V} \right\}.
\label{Eq:86}
\end{equation}
The connectivity between the thermodynamic description and the reconstructed $f(Q,T)$ gravity is given by Eqs.
\eqref{Eq:84}-\eqref{Eq:86}, where the total cosmic fluid are rewritten in terms of the effective quantities $\rho_{\rm DE}^{(f(Q,T))}$ and $p_{\rm DE}^{(f(Q,T))}$, which are determined by the modified Friedmann equations. At the background level these effective quantities are consisent with Eqs.~\eqref{eq:einstein1} and \eqref{eq:einstein2}, which ensure that the total density and pressure reduce to $\rho_{\rm tot}=3H^{2}/(8\pi)$ and $p_{\rm tot}=-(2\dot{H}+3H^{2})/(8\pi)$, respectively. Thus, Eq. \eqref{Eq:86} is consistent with and represented by the compact expression for the total entropy variation in Eq.~\eqref{Eq:81}. Then, the thermodynamic analysis is performed for the effective cosmic sector generated from the reconstructed $f(Q,T)$ theory, as made clear in Eqs.~\eqref{Eq:84}-\eqref{Eq:86}, and Eq.~\eqref{Eq:81} provides its convenient background-level form for the numerical evaluation. The resulting evolution of the total entropy variation is shown in Fig.~\ref{fig:entropy_fQT_consistency}. The compact background and effective-fluid descriptions exhibit the same behaviour, while the small residual shown in panel (c) quantifies their numerical consistency.

For the numerical evaluation, as pictorially presented in Fig. \ref{fig:entropy_fQT_consistency} of the entropy evolution, we first employ the CC-constrained power-law background, $H(z) = H_{0}(1+z)^{1/n}$, together with the Nojiri- Odintsov horizon entropy given by Eq.~\eqref{eq:SNO} and its time derivative in Eq.~\eqref{eq:SdotNO}. The matter contribution to the entropy is derived from the Gibbs relation in Eq.~\eqref{Eq:gibbs}, using the total effective cosmic fluid defined in Eq.~\eqref{Eq:84}. Here, the effective Dark Energy density and pressure are generated by the reconstructed $f(Q,T)$ dynamics, as explicitly given by Eqs.~\eqref{eq:rhoDE} and~\eqref{eq:pDE}. The resulting total entropy variation, obtained from Eq.~\eqref{Eq:86}, is then evaluated over the redshift interval $0 \le z \le 1.965$ using the CC-constrained background parameters alongside the reconstructed Nojiri-Odintsov parameters. For comparison, the same quantity is also evaluated directly from the compact background relations in Eq.~\eqref{Eq:81}. The two descriptions are subsequently compared through their numerical evolution and their corresponding consistency residual. The residual $\Delta\dot{S}_{\mathrm{tot}} = \dot{S}_{\mathrm{tot}}^{(f(Q,T))} - \dot{S}_{\mathrm{tot}}^{\mathrm{bg}}$, as seen in the third panel of Fig. \ref{fig:entropy_fQT_consistency}, is found to stay in the neighbourhood of zero over the redshift interval shown here. This indicates a quantitative consistency between the effective-fluid description associated with the reconstructed $f(Q,T)$ model and its compact background representation. The small deviation near the present epoch is attributed to numerical differentiation effects rather than an independent thermodynamic contribution. Overall, from Fig.\ref{fig:entropy_fQT_consistency} we find that the total entropy variation remains positive and evolves smoothly over the considered redshift range indicating that the generalized second law of thermodynamics is satisfied in the present framework. The good matching between the compact background and effective-fluid descriptions also indicates the consistency of the thermodynamic behaviour with the reconstructed $f(Q,T)$ dynamics, while the small residual is mainly a numerical consistency effect.


\section{Conclusions} \label{sec:conclusion}
In this study, we have examined the cosmological consequences of a modified gravity framework defined by $f(Q,T)= f(Q) + \lambda T$ in which a direct coupling between matter and geometry is introduced by the trace of the energy- momentum tensor. The required modified Friedmann equations were developed in terms of effective dark energy density and pressure by taking into account a spatially flat FLRW background. Using the effective equation of state parameter and the squared speed of sound, this formulation enabled us to investigate the cosmological behaviour of the modified gravitational sector. The study demonstrates that the effective cosmic dynamics is directly influenced by the matter-geometry interaction through $\lambda T$. To investigate the cosmological behaviour of non- linear $f(Q)$ corrections, we incorporated the phenomenological polynomial form, $f(Q)= Q+\alpha Q^m+\beta Q^q$, with $m>1$ and $q>m$, and a power-law scale factor $a(t)=a_0t^n$. The expressions for $f_Q, f_{QQ}$ were obtained and the effective Equation of state parameter and squared speed of sound were calculated. The power law cosmic background was constrained using the 32 cosmic chronometer readings in the redshift range of $0.07\leq z\leq1.965$. The $\chi^2$ yields the best-fit values $H_0= 64.0295~{\rm km s^{-1},Mpc^{-1}}$ and $n=1.02782$. The theoretical expansion history provides a satisfactory description of the CC data since the normalised residuals show no major systematic deviation, and the confidence contours shows positive correlation between $H_0$ and $n$. The best-fit values are within the acceptable confidence interval, indicating the validity of the power-law background.  Using these background parameters, we analyze the evolution of the effective Dark Energy equation-of-state parameter for some feasible parameter set. The resulting $w_{DE}$ remains finite and somewhat negative for the considered redshift range. In addition, the squared speed of sound remains positive throughout the redshift range showing that the effective dark energy sector is classically stable. Next, the GRGB-HDE sector with the logarithmically corrected horizon entropy, motivated by the generalized entropy framework of Nojiri, Odintsov and Faraoni \cite{Nojiri2022} was introduced in the $f(Q,T)$ framework by considering a correspondence between the holographic dark energy density and the effective dark energy density generated by the modified Friedmann equations. From this correspondence we had a first order linear differential equation for the geometric function $f(Q)$ that we have solved numerically without imposing any particular functional form for the reconstructed gravitational sector. Then the complete modified gravity function was obtained in the form of $f(Q,T)= f(Q)+\lambda T$ that preserves the coupling between matter and geometry. The reconstructed model is obtained for the optimised GRGB-HDE parameters with background parameters constrained by CC measurements and obtained as $H_0= 64.0295~{\rm km s^{-1},Mpc^{-1}}$ and $n=1.02782$. It is observed that the reconstructed geometric function has monotone increasing behaviour with $Q$, while the trace contribution $\lambda T$ stays negative. This implies that the observed decaying pattern of the full $f(Q,T)$ function is due to the enhanced importance of the trace contribution along the cosmological trajectory. From this reconstruction we can connect the generalized holographic dark energy directly to the underlying modified gravity dynamics. 

Finally, we used the Nojiri-Odintsov entropy with logarithmic correction to study the thermodynamic behaviour of the reconstructed $f(Q,T)$ model at the apparent horizon in order to further assess its thermodynamic viability. Using the Gibbs relation to combine the entropy of the horizon with the entropy of the effective fluid we find that the total entropy variation is positive for the range of redshifts considered here. This is in agreement with the generalized second law of thermodynamics. Furthermore, the evolutionary pattern of the entropy from the compact background description is consistent with that derived from the reconstructed $f(Q,T)$ effective fluid formulation. It is observed that the corresponding residual is small from which we understand that the two thermodynamic descriptions are mutually consistent except for some minute numerical differences resulting from the numerical evaluation of the reconstructed quantities. 

We refer to the recently established correspondence between generalized entropic cosmology and $f(Q)$ gravity \cite{NojiriOdintsov2025FQ,NojiriOdintsov2026FQ}. An important theoretical implication of the present reconstruction approch reported in this study is its connection with \cite{NojiriOdintsov2025FQ,NojiriOdintsov2026FQ}. The reconstruction method used here is holographically fixed by the generalized entropy prescription of $f(Q)$ and thus can be interpreted in the broader generalized entropic-$f(Q)$ setting. The additional term containing $\lambda T$ makes this geometric description a matter-geometry coupled $f(Q,T)$ theory. This gives a connection between the entropy based cosmological descriptions, holographic Dark Energy and the non-metric modified gravity.

\section*{Data Availability}

The observational datasets used in this work are taken from \cite{Favale2023} duly acknowledged in the bibliography.

\section*{Acknowledgements}

The visiting associateship of the Inter-University Centre for Astronmy and Astrophysics (IUCAA), Pune, India, is thankfully acknowledged by Surajit Chattopadhyay.

\end{document}